\documentclass[twocolumn,prl,aps]{revtex4}
\usepackage{graphicx}

\begin{document}

\title{Do Hardy-type proposals test nonlocality of single-particle?}

\author{Antoine Suarez}
\address{Center for Quantum Philosophy, CH-Berninastrasse 85, 8057 Zurich, Switzerland\\
suarez@leman.ch, www.quantumphil.org}

\date{April 15, 2013}

\begin{abstract}

It is argued that Hardy-type proposals assume either nonlocality at detection, and then beg the question, or ``local empty waves'', and then have to accept ``many worlds'' and cannot prove nonlocality.

\end{abstract}


\maketitle

\noindent \textbf{Introduction}.\textemdash In the 5th Solvay conference in 1927 Einstein presented a simple single-particle gedanken-experiment showing that the idea of the quantum mechanical collapse of the wave function implies nonlocal actio at a distance and therefore conflicts with relativity. However after the Solvay conference Einstein withdrew from this argument to the more complicated EPR one he published in 1935.\cite{bv}

The original 1927 Einstein's experiment has been recently done for the first time. The results demonstrate single-photon nonlocality at detection and in addition highlight something Einstein did not explicitly mention: Nonlocality at detection is necessary to have conservation of energy in each single quantum event \cite{gszgs, as12}. In other words, on the basis of his 1927 argument Einstein could not object to quantum mechanics without rejecting the conservation of energy. By contrast the EPR argument (1935) allowed him to argue against quantum mechanics without giving up the conservation of energy.

In this Letter I compare our experiment \cite{gszgs, as12} with the probe of single-particle nonlocality proposed by Lucien Hardy in 1994 \cite{lh}, and improved thereafter by other authors \cite{dv,hv}. I show that Hardy's proposal takes clearly for granted the result of the experiment presented in \cite{gszgs, as12}. Therefore Hardy-type proposals either tacitly assume as an axiom the nonlocality they aim to prove, or they assume the de Broglie's ``particle and empty wave'' as local hidden variables. This means that Hardy-type proposals rule out such local hidden variables (although contrarily to Bell-type experiments without invoking inequalities), but do not prove nonlocality.
\\

\begin{figure}[t]
\includegraphics[width=60 mm]{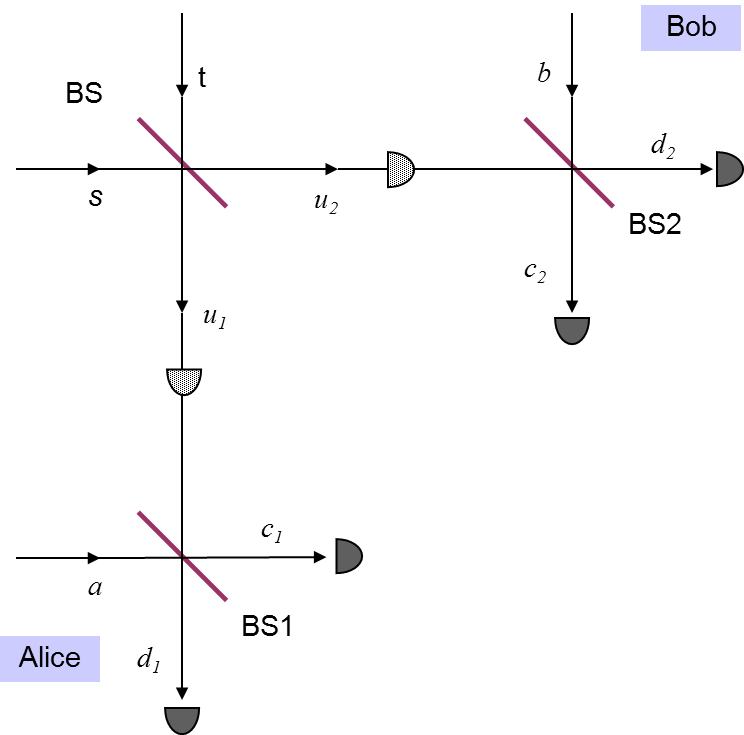}
\caption{Hardy's scheme: Alice/Bob can either directly set a detector on her/his path (dashed detectors), or she/he can make a homodyne detection by combining her/his path with a local oscillator at a 50:50 beam splitter BS1 respectively BS2 (solid detectors). The source (not represented) is supposed to be single-photon. The vacuum is incident on the mode $t$; the input state is incident on mode $s$ with no more than one photon at any time; the coherent states incident on modes $a$ and $b$ act as local oscillators.}
\label{f5}
\end{figure}

\noindent \textbf{Hardy's proposal}.\textemdash The argument is conceived as a ladder of four gedanken-experiments using the scheme sketched in Figure \ref{f5} \cite{lh,dv}. Alice and Bob each have two choices, which lead to the four possible \emph{Experiments 1-4} described below. The input state incident on $s$, and the corresponding coherent states incident on $a$ and $b$ are chosen in a way that quantum mechanics yields the outcome predictions correspondingly listed to the following \emph{Experiments 1-4}:
\\

\emph{Experiment 1}: Alice and Bob both decide to set a detector on their paths $u_1$ respectively $u_2$ (dashed detectors).

\emph{Quantum mechanical prediction}: If Alice detects a particle on $u_1$, then she can infer that Bob detects nothing on $u_2$, and viceversa, if Bob detects a particle on $u_2$, he can infer that Alice detects nothing on $u_1$. (Notice that if Alice (Bob) detects nothing on $u_1$ ($u_2$), she (he) cannot infer that Bob (Alice) detects a particle on $u_2$ ($u_1$)).

Regarding this \emph{Experiment 1} Hardy's argument states: ``In this case, it is clear that they cannot both detect a photon since no more than one photon is emitted from the source at any time. This means that detecting a particle on $u_1$ and detecting a particle on $u_2$ never happens.'' \cite{lh,dv}
\\

\emph{Experiment 2}: Alice makes a homodyne detection at $c_1$ and $d_1$, by combining the state on path $u_1$ with a local oscillator (coherent state) $a$, at her 50:50 beam splitter BS1. Bob, meanwhile, makes the same measurement as in \emph{Experiment 1}.

\emph{Quantum mechanical prediction}: If Alice detects a particle at $d_1$ and nothing at $c_1$, then Bob must detect one particle.
\\

\emph{Experiment 3}: The roles of Alice and Bob are reversed: Alice sets her detector on path $u_1$ and Bob makes a homodyne detection at $c_2$ and $d_2$.

\emph{Quantum mechanical prediction}: if Bob detects one particle at $d_2$ and nothing at $c_2$, then he can infer that Alice must have detected a particle in path $u_1$.
\\

\emph{Experiment 4}: Both Alice and Bob choose to make homodyne detections.

\emph{Quantum mechanical prediction}: According to quantum mechanics the experiment can have different outcomes. One of the possible ones is that Alice records one particle at $d_1$ and nothing at $c_1$, while Bob records one particle at $d_2$ and nothing at $c_2$.
\\

Suppose now the following reasoning: The result of \emph{Experiment 4} contradicts \emph{Experiment 1}: Indeed, according to \emph{Experiment 2} Alice infers from her measurement that a single particle must have traveled along path $u_2$ towards Bob. At least, this is true in the sense that, had Bob put his detector in path $u_2$, he would have been guaranteed to detect a particle. However, according to \emph{Experiment 3}, at the same time Bob infers from his measurement that a single particle must have traveled along path $u_1$ towards Alice. According to \emph{Experiment 1} Alice and Bob cannot both be right.
\\

Hardy pointed out that this reasoning does tacitly assume locality, and without this assumption the contradiction disappears: Alice might deduce from her result that, had Bob measured the number of particles in path $u_2$, he would definitely have detected exactly one. However, if Bob had measured $u_2$ instead of the homodyne measurement he did make, there might have been a nonlocal influence from Bob's end to Alice's end and then she might have obtained a different measurement outcome.\cite{lh,dv}
\\

\noindent \textbf{Hardy takes for granted the result of \cite{gszgs}}.\textemdash The first step of Hardy's ladder,i.e., \emph{Experiment 1}, has clearly the character of an axiom stating the principle of ``one photon one count'', i.e., conservation of energy in each individual quantum event, in the very same experiment presented in \cite{gszgs}. If one assumes that in this experiment the outcome becomes determined at the detection \cite{as12}, then Hardy's axiom \emph{Experiment 1} would tacitly assume the nonlocality at detection demonstrated in \cite{gszgs}, and Hardy's argument would assume just the nonlocality it searches to prove.
\\

\noindent \textbf{Hardy-type experiments refute the de Broglie's local ``empty wave'' model}.\textemdash Consequently to avoid begging the question Hardy-type experiments have to assume that the outcome of the experiment in \cite{gszgs} is decided at the beam-splitter, and this means to accept de Broglie's ``local empty wave'' in order to be able of explaining the quantum mechanical predictions in interference experiments \cite{as12}. Consider the statement at the core of Hardy's argument:

``if Bob had measured $u_2$ instead of the homodyne measurement he did make, there might have been a nonlocal influence from Bob's end to Alice's end and then she might have obtained a different measurement outcome''.

This amounts to explain things as follows in terms of decisions at the beam-splitters and local ``empty waves'':

Suppose the empty wave travels along path $u_1$ and the particle along $u_2$. If these paths are not monitored, nonlocal coordination between the decisions at the beam-splitters BS1 and BS2 is possible and the result expected in \emph{Experiment 4} may appear. By contrast if Alice had put her detector on $u_1$, the ``empty wave'' traveling this path would not have reached BS1 and therefore the nonlocal coordination between the decisions at BS1 and BS2 would disappear. Consequently Bob would obtain a different result, that is one that does not allow him to conclude there is a particle on $u_1$.

In other words, what Hardy-type proposals test are models assuming outcome's decision at the beam-splitters with the particle leaving by one output port and the corresponding ``empty wave'' by the alternative output port.\cite{jb} This means that a Hardy-type experiment would refute such ``local hidden variables'' (particle and empty wave) not through an statistical result (like the conventional Bell-type experiments do), but through a single measurement like the GHZ argument does \cite{ghzz}.

However, as we will see in the following section, de Broglie's local ``empty wave'' model can be straightforwardly completed with additional particles and ``empty waves'' to a deterministic and local version of the ``many worlds'' interpretation of quantum mechanics called ``parallel lives'', which is not refuted by either Hardy's argument or violation of Bell's inequalities.\cite{brr}
\\

\noindent \textbf{From the local ``empty wave'' to local ``parallel lives''}.\textemdash ``Empty waves'' are entities that exist and propagate within space-time but neither carry energy nor momentum, and are not directly accessible to general observation. ``Empty waves'' interact with the environment only in a very selective and specific way - actually an ``empty wave'' does not interact with any particle (and for this reason cannot be detected), but only with``its particle''.

If one accepts that at beam-splitter BS a particle P splits into a transmitted particle P' (leaving by output-port T) and a reflected ``empty wave'' W' (leaving by output-port R), one can as well assume that the split produces additionally the alternative outcome, that is, a second transmitted empty wave W* (leaving by output-port T) and a reflected particle P* (leaving by output-port R) under the following conditions: The particle\&wave pair (P'\&W') lives with a copy of Alice within a space-time ``bubble'' denoted Alice', and the pair (P*\&W*) lives with another copy of Alice within a parallel ``bubble'' denoted Alice*; the ``bubbles'' Alice' and Alice* cannot interact with each other; however in 2-particles entanglement experiments the bubble Alice' - containing the outcome (P'\&W')$_A$ - may interact with (depending on the experiment) either the bubble Bob' -containing the outcome (P'\&W')$_B$ - or the bubble Bob* - containing the outcome (P*\&W*)$_B$, and this way account for the characteristic quantum (EPR) correlations.

This is basically the explanation provided by the ``parallel lives'' interpretation, a version of ``many worlds'' recently proposed by Gilles Brassard and Paul Raymond-Robichaud \cite{brr}. Both ``empty waves'' and ``parallel lives'' are based on the assumption that within space-time one can have entities or regions that interact with the environment only in a selective and predetermined way.

The analysis by Brassard and Raymond-Robichaud shows the importance of the following \emph{Principle}:

\emph{All that is in space-time is accessible to observation (except in the case of space-like separation).}

If you are going to be consistent, you are only capable of opposing the ``many worlds'' and ``parallel lives'' interpretations if you accept this \emph{Principle}.
\\

\noindent \textbf{Refutation of the local ``empty wave'' doesn't mean refutation of locality}.\textemdash``Parallel lives'' accounts for a unitary evolution of the quantum state and for the quantum correlations invoking exclusively local causality, i.e. causal links within the light cone.\cite{brr} Therefore ``parallel lives'' shows that the locality lost through refutation of local models based on outcomes consisting of \emph{one couple} ``particle \& empty wave'' leaving the beam-splitter, can easily be restored by simply improving the models with outcomes consisting of \emph{two couples} ``particle \& empty wave''. In other words, neither Hardy's argument nor violation of Bell's inequalities rule out the local and deterministic model of ``parallel lives''.

``Many worlds'' was originally formulated by Hugh Everett in 1957 as an attempt to overcome quantum mechanical paradoxes like the ``Schr\"{o}dinger cat'' derived from the ``measurement problem''. Nevertheless John Bell suggested that the picture may have to do with nonlocality as well:

``The 'many world interpretation' seems to me an extravagant, and above all an extravagantly vague, hypothesis. I could also dismiss it as silly. And yet... It may have something distinctive to say in connection with the 'Einstein Podolsky Rosen puzzle', and it would be worthwhile, I think, to formulate some precise version of it to see if this is really so.'' (\cite{jb} p. 194).

Work by Lev Vaidman \cite{lev}, and the more recent ``parallel lives'' interpretation by Gilles Brassard and Paul Raymond-Robichaud \cite{brr}, show that ``many worlds'' has really something distinctive to say in connection with the ``EPR puzzle'': If one accepts de Broglie's ``empty wave'' it is impossible to oppose the ``many worlds'', and therefore, refutation of de Broglie's local hidden variables model (be it by Hardy's argument or the violation of Bell's inequalities) doesn't mean refutation of locality.

There is however a crucial difference between Hardy's argument and Bell's one. If one assumes that the outcome of a Hardy-type experiment (Figure \ref{f5}) becomes decided at detection (and therefore one disposes of the local ``empty wave''), Hardy's argument cannot prove nonlocality without assuming the result of \cite{gszgs} as an axiom (``Experiment 1''), and therefore begs the question. By contrast the experimental violation of Bell's inequalities refutes locality independently of the result of \cite{gszgs}. Nonetheless for the time being Bell-type experiments exhibit the detection loophole, and hence strictly speaking the only to date loophole-free demonstration of nonlocality is the presented in \cite{gszgs}. In this context it is noteworthy as well that the only QKD available on the market to date is based on the BB84 protocol, and this means on the nonlocality proved in \cite{gszgs}.

Whether Hardy-type experiments can be considered ``single-particle'' ones is matter of debate \cite{ghz,dv} (likely because the concept of ``single-particle experiment'' has not been well defined). As a matter of fact Hardy-type experiments use 4 detectors, like the Bell-type experiments (instead of only 2, like the experiment presented in \cite{gszgs, as12}). But in any case this issue is not of interest for the sake of our discussion.
\\

\noindent \textbf{Conclusion}.\textemdash In conclusion, Hardy-type experiments either assume the nonlocality they aim to prove, or assume the ``local empty wave'' and don't prove nonlocality. By contrast ``[the] nonlocality of a single particle expressed by Einstein at the 1927 Solvay Conference'' \cite{lh,dv} is demonstrated by the experiment presented in \cite{gszgs}. The nonlocality this experiment proves \cite{as12} is more basic than Bell's nonlocality and rules the whole quantum mechanics: It appears already at the level of single particle experiments (requiring only 2 detectors), and not first in entanglement experiments involving 2 or more particles (requiring at least 4 detectors). This may open the road to deriving the ``abstract quantum algebra'' from principles with a clear physical content like free will, conservation of energy, and locality emergent from nonlocality \cite{as10a}.

\emph{Acknowledgments}: I am thankful to Antonio Acin, Gilles Brassard, Thiago Gerreiro, Nicolas Gisin, Stefano Pironio, Bruno Sanguinetti, and Hugo Zbinden for stimulating discussions.


\begin{references}

\bibitem{bv} Guido Bacciagaluppi, Antony Valentini, Quantum Theory at the Crossroads: Reconsidering the 1927 Solvay Conference. Part III: The proceedings of the 1927 Solvay conference. Cambridge University Press, 2009. arXiv:quant-ph/0609184v2 (2009)

\bibitem{gszgs} T. Guerreiro, B. Sanguinetti, H. Zbinden, N. Gisin and A. Suarez, Single-photon space-like antibunching. \emph{Phys Let A} 376 (2012) 2174-2177. http://dx.doi.org/10.1016/j.physleta.2012.05.019. arXiv:1204.1712v1 (2012).

\bibitem{as12} A. Suarez, `Empty waves'', ``many worlds'', ``parallel lives'', and nonlocal decision at detection, arXiv:1204.1732 (2012); see also A. Suarez and P. Adams (Eds.) Is Science compatible with free will? Exploring free will and consciousness in light of quantum physics and neuroscience. Springer, New York, 2012, Chapter 5.

\bibitem{lh}  L. Hardy, Nonlocality of a single photon revisited, \emph{Phys. Rev. Let.} 73,(1994) 2280

\bibitem{dv} J. Dunningham and V. Vedral, Nonlocality of a single particle, \emph{Phys. Rev. Let.} 99, (2007) 180404-2

\bibitem{hv} L. Heaney, A. Cabello, M. Fran\c{c}a Santos, and V. Vedral, Extreme nonlocality with one photon, arXiv:0911.0770v3 (2011).

\bibitem{ghzz} D. Greenberger, M. Horne, A. Zeilinger and M. Zukowski, A Bell Theorem Without Inequalities for Two Particles, Using Inefficient Detectors, 	arXiv:quant-ph/0510207v2 (2008).

\bibitem{jb} J.S. Bell, {\em Speakable and unspeakable in quantum mechanics}, Cambridge: University Press, 1987.

\bibitem{lev} Lev Vaidman, The Many-Worlds Interpretation of Quantum Mechanics. \emph{The Stanford Encyclopedia of Philosophy} (Summer 2002 Edition), E. N. Zalta (ed.)

\bibitem{brr} G. Brassard and P. Raymond-Robichaud, Can Free Will Emerge from Determinism in Quantum Theory? In: A. Suarez and P. Adams (Eds.) Is Science compatible with free will? Exploring free will and consciousness in light of quantum physics and neuroscience. Springer, New York, 2013, Chapter 4. arXiv:1204.2128v1.

\bibitem{ghz} D. Greenberger, M. Horne, A. Zeilinger, Nonlocality of a single photon? \emph{Phys. Rev. Let.} 75, (1995) 2064

\bibitem{bb84} C.H. Bennett and G. Brassard, Quantum Cryptography: Public Key Distribution and Coin Tossing, in: \emph{Proceedings of IEEE International Conference on Computers Systems and Signal Processing}, Bangalore India (1984) 175-179.

\bibitem{as10a} A. Suarez, Deriving Bell's nonlocality from nonlocality at detection. arXiv:1009.0698 (2010).

\end{references}
\end{document}